\documentclass[final]{raa}   
\usepackage{graphicx,times}
\usepackage{natbib}
\usepackage{amssymb,amsmath}
\usepackage{siunitx}
\usepackage{textcomp}
\bibpunct{(}{)}{;}{a}{}{,}

\usepackage{longtable}

\newcommand{\insight}{{\it Insight-HXMT}}

\begin{document}

   \title{\insight{} observations of the Crab pulsar}

 \volnopage{ {\bf 20XX} Vol.\ {\bf X} No. {\bf XX},000--000}
   \setcounter{page}{1}

   \author{Youli Tuo
      \inst{1,2}
   \and Mingyu Ge
      \inst{1}
	\and Liming Song
      \inst{1}
    \and Linli Yan 
      \inst{3}	
      \and Qingcui Bu
      \inst{1}
      \and Jinlu Qu
      \inst{1}
   }

   \institute{Key Laboratory of Particle Astrophysics, Institute of High Energy Physics, Chinese Academy of Sciences, Beijing 100049, China; {\it tuoyl@ihep.ac.cn}\\
        \and
              University of the Chinese Academy of Sciences, Beijing, Beijing 100049, China\\
              \and 
              School of Mathematics and Physics, Anhui Jianzhu University, Hefei 230601, China \\
\vs \no
   {\small Received 20XX Month Day; accepted 20XX Month Day}
}

\abstract{We report on X-ray emission properties of the Crab pulsar(PSR B0531+21) using observations by \insight{} during its first year after launch. We obtained high signal-to-noise profiles in the X-ray energy band 11--250\,keV. We have confirmed an increase in the flux ratio of the second peak over the main peak with increasing energy, consistent with other missions. The separation of the two peaks shows no significant trend with increasing energy. 
The phase-averaged spectrum, fitted by a {\it logpar} model, and the phase-resolved spectra of the Crab pulsar, fitted by a {\it powerlaw} in the different energy bands of {\it HXMT}, are consistent with {\it RXTE} and {\it NuSTAR} in that photon indices evolve as a function of phase as well as a function of energy contributing to a broadband modeling.
}

   \authorrunning{Youli Tuo, Mingyu Ge \& Liming Song et al. }            
   \titlerunning{\insight{} observation of the Crab pulsar}  
   \maketitle

%
\section{Introduction}\label{sec:intro}

The Crab pulsar(PSR B0531+21), a central energetic  engine for the Crab Nebula, is  a young archetypical isolated rotation-powered  pulsar with rotational energy $\dot{E} \approx 5 \times 10^{38}\,\mathrm{erg}\,\mathrm{s}^{-1}$ which has been studied sell in almost the entire electromagnetic spectrum. Because of its relatively stable evolution in period($P = 33.6\,\mathrm{ms}$) with its derivative $\dot{P} \approx 4.2 \times 10^{-13}\, \mathrm{s}\,\mathrm{s}^{-1}$ and brightness, it has been widely used as a calibration source for astronomical missions. In the X-ray energy band, missions including {\it BeppoSAX}, {\it RXTE}, {\it Chandra}, {\it INTEGRAL}, and {\it NuSTAR} have provided broadband observations of the Crab pulsar and nebula\citep{Massaro2000, Rots2004, weisskopf2004chandra, Mineo2006, madsen2015broadband}, the results of which suggested a power-law shaped phase-averaged spectrum for the nebula together with the Crab pulsar. The photon index of the Crab pulsar in the soft X-ray band fitted by a power-law model was 2.022\citep{Kuiper2001}. The Crab Nebula + pulsar in the 1--100\,keV range was fitted by a power-law model with photon index of 2.1({\it RXTE}, {\it BeppoSAX}, {\it INTEGRAL}/JEM\_X)\citep{kirsch2005crab}, while {\it NuSTAR} results covering 3--79\,keV agreed on a photon index Γ = 2.1 within circles of radius $200''$ \citep{madsen2015broadband}. Above 100\,keV, hard X-ray results of {\it INTEGRAL} suggested a softer index of $\Gamma \sim$ 2.20-–2.25. It is a calibration source for \insight{} as well, so results on the Crab pulsar indicate the performance of the instruments.
\par
Multiple models were employed to interpret the observations during the last 50 years(see recent reviews on pulsars, e.g., \cite{harding2017pulsar,	Venter2018}). The polar cap model suggests that the $\gamma$-ray emission originates from curvature radiation of electrons accelerated at the polar caps moving along the magnetic field lines. 
\cite{Cheng1985} points out that particles could be produced and accelerated by gaps which are located between the null charge surface and light cylinder. Those particles emitting curvature radiation in $\gamma$-rays could be supplied by the photon-photon process using thermal X-rays from the neutron star surface. Since emission models yielded different regions for accelerating particles, phase-resolved spectroscopy is essential for constraining these models and provides comprehensive details about different regions of a pulsar as it rotates.

 In this paper,  we present the \insight{} data and conduct timing and phase-resolved spectral analysis. X-ray properties of the Crab pulsar in the \insight{} energy band are introduced. The data reduction is described in Section \ref{sec:obsandred}. The data analysis applied to timing and spectral properties is presented in section \ref{sec:data analysis}. In section \ref{sec:results}, we provide the results on timing and spectral analysis including the properties of pulse profiles and spectra.
In Section \ref{sec:disscussion}, the main conclusions are discussed.

\section{Observations and data reduction}\label{sec:obsandred}
The {\it Hard X-ray Modulation Telescope}({\it HXMT}), dubbed \insight{}, was launched on 2017 June 15. There are three slat-collimated instruments\citep{zhang2014introduction}, the High Energy X-ray telescope(HE, 20--250\,keV, 5100\,$\mathrm{cm}^2$) , the Medium Energy X-ray telescope(ME, 5--30\,keV, 952\,$\mathrm{cm}^2$) and the Low Energy X-ray telescope(LE, 1--15\,keV, 384\,$\mathrm{cm}^2$) enabling research on broadband X-ray spectroscopy. We employed the HE and ME instruments to investigate the timing properties of the Crab pulsar because of their relatively good time resolution($\sim$2\,\textmu{}$\mathrm{s}$ for HE, $\sim$20\,\textmu{}$\mathrm{s}$ for ME). The energy range selected for ME and HE is 11--27\,keV and 27--250\,keV respectively. There are three types of field of view (FoV): $1^\circ \times 6^\circ$(full-width at half-maximum(FWHM), also called the small FoV),$6^\circ\times6^\circ$(the large FoV), and the blind FoV. We selected the photons observed by small FoV detectors for ME because the large FoV was easily contaminated by a nearby source and the bright Earth.\citep{Chen2018,Huang2018}

Observations of the Crab pulsar were started on2017 August 27. The data in time range MJD 57992--58209 were employed in this work. 
The exposure times for each observation ID are listed in Table \ref{table:exposure}. The pointed observation mode was selected for the following analysis.

\begin{table}[ht]
\caption{ \insight{} pointed observation Mode Targeting the Crab pulsar}\label{table:exposure}

\centering
\begin{tabular}{lcllr}
\hline\hline
Instrument  & Obs ID & start MJD  &  stop MJD  &  Exposure(ks)\footnote{exposure time after the data reduction}                \\ 
\hline

ME(\si{5}--\SI{30}{\keV})  & P0101299 & 57992.2 & 58027 & 158.3 \\
ME(\si{5}--\SI{30}{\keV}) & P0111605 & 55927.0 & 58209.3 & 296.8 \\

HE(\si{20}--\SI{250}{\keV}) & P0101299 & 57992.2 & 58026.8 & 156.3 \\
HE(\si{20}--\SI{250}{\keV})  & P0111605 & 58066.2 & 58205.8 & 411.7 \\
\hline

\end{tabular}
\end{table}

The data reduction process was conducted via {\it HXMTsoft(v2.0)}\footnote{http://www.hxmt.org/index.php/dataan} as follows. For the HE instrument, photons recorded by the NaI detector on NaI/CsI  Phoswich modules with a pulse width between 54 and 70 were selected. We identified good time intervals for HE by applying theses criteria: (1) $ELV$(the elevation of point source above the horizon) larger than $6^\circ$, (2) $COR$ (the geometric cutoff rigidity) larger than 8, (3) $ANG\_DIST$(offset angle from the pointing direction to coordinate of the Crab pulsar) smaller than $0.1^{\circ}$, and (4) time before and after South Atlantic Anomaly(SAA) region, $T\_SAA$ and $TN\_SAA$ bigger than \SI{100}{\second}. The criteria for good time intervals applied to ME are $ELV\textgreater 6^\circ$, $COR\textgreater10$, $ANG\_DIST\textless0.1^\circ$, $T\_SAA\textgreater100\mathrm\,{s}$, and $TN\_SAA\textgreater100\,\mathrm{s}$. The good time intervals for the two instruments are slightly different from each other due to different field of FOVs, on which the background signals have distinct effects, despite that sources were originally observed simultaneously by different instruments. The time of arrival(ToA) for photons detected by the two instruments was converted to ToA to Solar System Barycenter using ephemeris DE405 by HXMTsoft {\it hxbary}.

\par To revmove the background signal from data prepared in the previously described way, a timing analysis strategy was employed in which we selected photons in a specific phase range during each pulsar period as background photons.

\section{Data analysis}
\label{sec:data analysis}

\subsection{timing analysis}
\label{subsec:timing analysis}

We processed the timing analysis as follows. First we calculated the periodic phase for each photon at the barycenter of the Solar System using parameters in the ephemeris derived from {\it HXMT} observation. The phase of each photon is calculated by

\begin{equation}\label{eq:Phi}
\phi = \nu(t-t_0) + \frac{1}{2}\nu_{1}(t-t_0)^2 + \frac{1}{6}\nu_{2}(t-t_0)^3 + \cdots,
\end{equation}

where $t_0$ is the reference time, $\nu$ is the periodic frequency, and $\nu_{1}$ $\nu_{2}$ are derivative  and second derivative of frequency for reference time, respectively. We normalized the value of phases using the decimal part of each phase. 
We searched the frequency  $\nu$ of the Crab pulsar every 10 hours after observations started by using the Pearson $\chi ^2$ test, and fitted the frequency with a polynomial function, where frequency evolves as $\Delta \nu = \nu_{1}(t-t_{0}) + \frac{1}{2}\nu_{2}(t-t_{0})^2 + \frac{1}{6}\nu_{3}(t-t_{0})^3 + \frac{1}{24}\nu_{4}(t-t_{0})^4 + \cdots$. The parameters $\nu$, $\nu_1$, $\nu_2$, $\nu_3$ and $\nu_4$ describing the spin could be obtained. Theses parameters are listed in Table \ref{table:para}, in which the epoch is the time range of spinning parameters that fits the profile. Due to the evolution of these parameters, the spinning parameters describe the Crab pulsar's spinning behavior within the time range of an epoch. 

 Since we obtained the profiles every 10 hours, we calculated the timing residuals of ToAs. ToA of a profile is defined as $t_0+\phi_{\mathrm{peak}}*P$, where $t_0$ is the reference time, $\phi_{\mathrm{peak}}$ is the phase of the main peak(MP) and $P$ is the period. Timing residuals are the differences between the calculated ToA of the MP and the observed ToA of the MP in each profile. The root mean square(rms) values of timing residuals are presented in Table \ref{table:para}.

\begin{table}[htbp]
\caption{Spinning parameters of the Crab pulsar}
\scriptsize{} \label{table:para}
\begin{center}
\begin{tabular}{ llllllc}
\hline\hline
&Epoch(MJD)                                      &     57929.54--58439.50        &   58066.17--58097.61          &   58097.60--58147.61         &   58147.60--58209.62\\
&$\nu$(Hz)                                        &     29.639428373(2)              &29.636679769(3)                    & 29.63668019(3)                    &     29.6334830468(4)      \\
& $\nu_1$\,(10$^{-10}$Hz\,s$^{-1}$) & -3.68719(3)                           &-3.6981(3)                               &  -3.7153(4)                            &     -3.692902(5)    \\
& $\nu_2$\,(10$^{-20}$\,Hz\,s$^{-2}$) & 3.5(3)                                     & -384(2)                                   &  66(4)                                      &      9.5(5) \\
& $\nu_3$\,(10$^{-27}$\,Hz\,s$^{-3}$) & -9(1)                                      & 8728(591)                               &  -176(25)                                 &      -30(3) \\
& $\nu_4$\,(10$^{-33}$\,Hz\,s$^{-4}$) & 0                                            & -11671(1108)                            &  31(6)                                         &      4(2) \\
& $\nu_5$\,(10$^{-36}$\,Hz\,s$^{-5}$) & 0                                            & 7.5(9)                                     &  0                                               &     0 \\
& $t_0$(MJD)                                   & 57979.425942180467246     &  58066.180875391473819   &  58066.180875391473819        &  58166.180875391473819 \\
& rms(\si{\us})                                 & 22.320                                   & 30.179                                    & 19.417                                      & 24.931  \\

\hline
\end{tabular}
\end{center}
\end{table}

\begin{figure}
	\centering
	\includegraphics[width=1.0\textwidth]{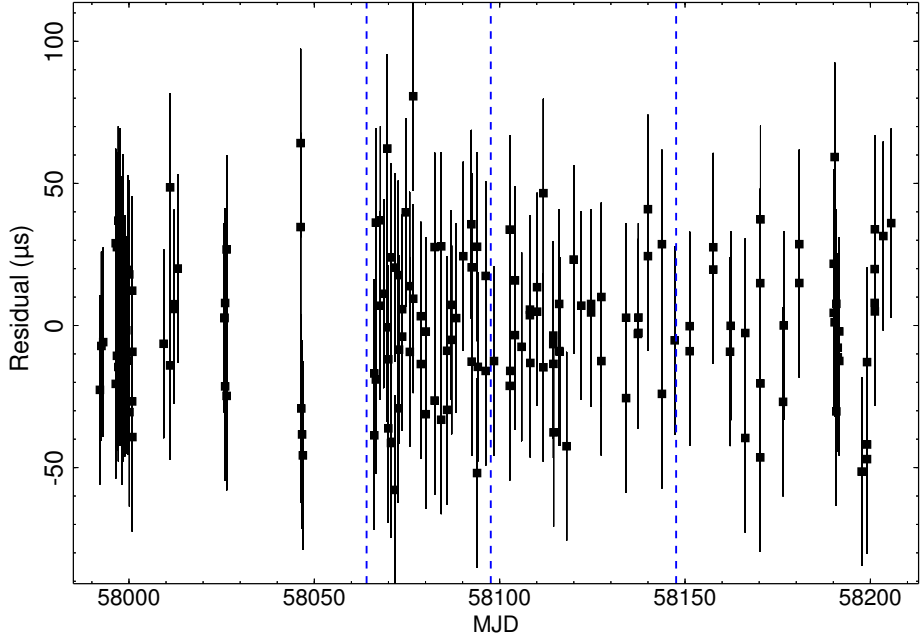}
	\caption{Timing residuals for four different epoch ranges as indicated by three blue dash line. Timing residuals for each time range were fitted with different ephemerides. 
	\label{fig:timing_resi}}
\end{figure}

In order to analyze the evolution of profiles in energy, the light curve profiles for all observations in different energy bands were generated using phases we obtained. We selected eight energy bands based on the statistical significance of each profile. They are 11-15\,keV, 15-20\,keV, 20-25\,keV, 25-30\,keV, 30-35\,keV, 35-50,keV, 50-100,keV and 100-250,keV. The MPs of profiles from different energy bands were aligned together by calculating the cross-correlation function. We set the phase of the MP maximum as phase 0.  The off-pulse phases, 0.6-0.8, which refer to phases of emission from the Crab Nebula and the surface of the pulsar, were chosen as background phases, an approach which has been applied in previous results for X-ray observations by {\it RXTE}-PCA($2--60\,\mathrm{keV}$)\citep{ge2012x, Yan2018}. Profiles were normalized as values of MP being 1, and the average values of off-pulse phases being 0.

\begin{figure}[!thb]
\centering
\includegraphics[width=1.0\textwidth]{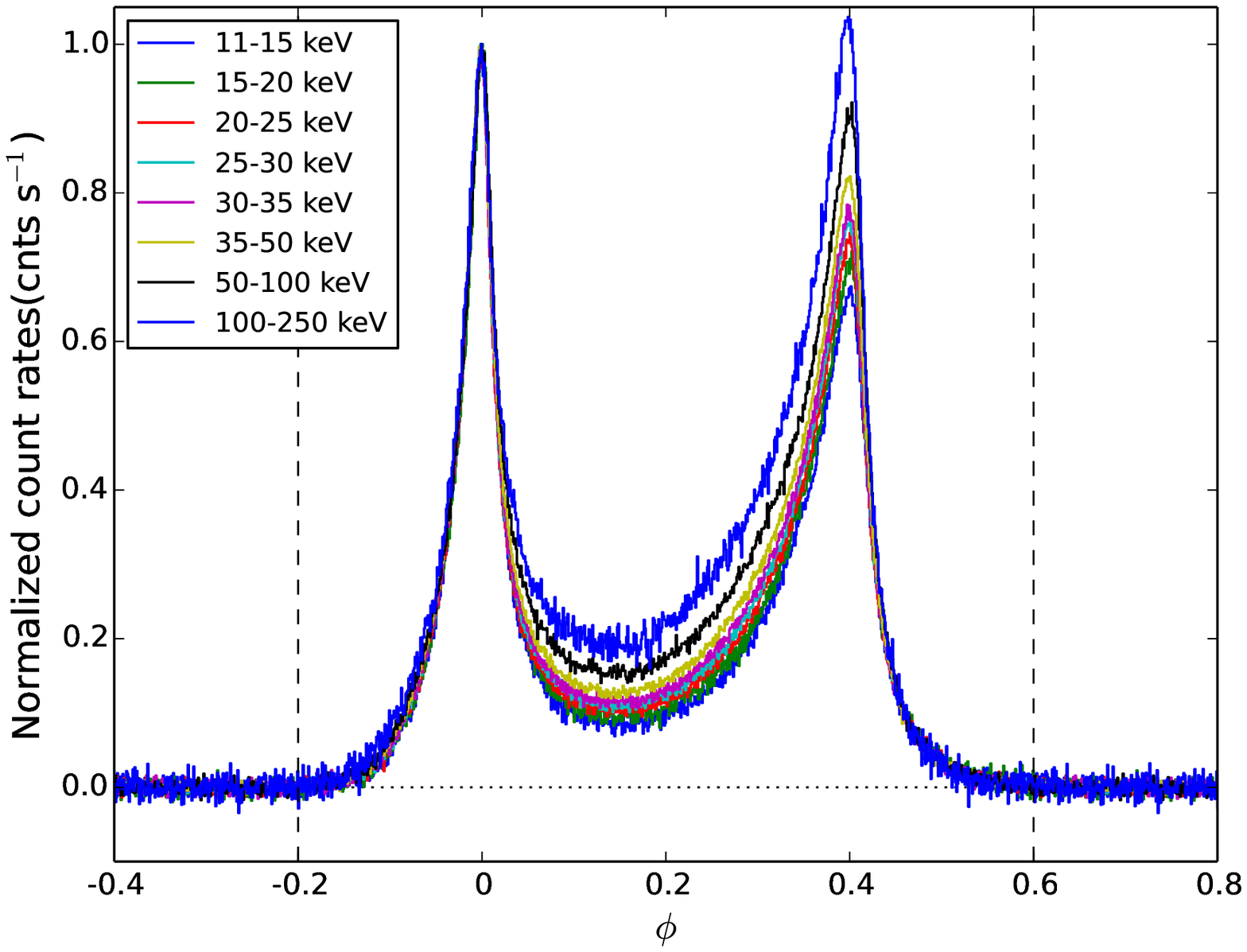}\caption{Pulse profiles for different energy bands covering HE and ME. The location of the maximum of MP is aligned as phase 0. The off-pulse region(phase 0.6-0.8) is selected as background. The count rate of profiles is normalized by subtracting the mean count rate of the off-pulse region and dividing by the difference of between the maximum count rate of MP and the mean count rate of background,  }\label{fig:profiles_energy}
\end{figure}

To obtain precise values between the phase separation($\Phi$) of two peaks, and FWHMs of thesepeaks($W_1$,$W_2$), we fitted the MPs and the secondary peaks (SPs) of profiles separately with an empirical formula \citep{nelson1970optical}

\begin{equation}\label{eq:profile_fitting}
	L(\phi) = N \frac{1+c_1(\phi -\phi_0) + c_2(\phi-\phi_0)^2}{1+c_3(\phi - \phi_0) + c_4(\phi - \phi_0)^2}e^{-c_5(\phi - \phi_0)^2} + l
\end{equation}
where the $L(\phi)$ is the normalized counts of the profile at phase $\phi$. Parameters $c_1$,$c_2$,$c_3$,$c_4$, and $c_5$ are the shape coefficients affecting the shape of the peak. Parameter $\phi_0$ shifts the profiles. $N$ is the pulse height of the profile, and $l$ is the baseline of the profile. Eq.\ref{eq:profile_fitting} was only applied to fitting peaks rather than whole profiles, thus only part of profiles, phases -0.04 to 0.03 and phases 0.33 to 0.42 are selected as MP and SP for fitting respectively. Then, the separation between the two peaks could be calculated by the fitted profiles, and the uncertainty is given by the variance of parameter $\phi$ which shifts the profiles in phase.

\subsection{spectral analysis}
\label{subsec:spec}

\par To analyze the spectral properties of the Crab pulsar, all spectra were fitted using {\it XSPEC} (Version 12.10.0c) \citep{arnaud1996xspec} with errors given at the $1.64\sigma$ (90\%) confidence level. We present the phase-averaged spectrum in energy range 11--250\,keV with a log parabola model({\it logpar}, Eq.\ref{eq:logpar}), which contributed to a linear dependence of the spectral slope upon the logarithm of the photon energy $E$ 
\begin{equation}\label{eq:logpar}
	F(E) = K(E/E_0)^{-[\alpha + \beta \mathrm{Log} (E/E_0)]} .
\end{equation}
For the parameters of {\it logpar} models with an interstellar absorption model, we froze the $N_H$ at 0.36 ($10^{22}$\,atoms\,cm$^{-2}$), while $\alpha$, $\beta$ and the normalization value were set as free parameters to fit. The background spectrum was extracted between phases 0.6 and 0.8. Since the background phases were 1/5 of all the phases, the exposure time of the background spectrum was 1/5 of the total exposure time. For  ME, all observed data of all small FoV detectors contributed to one spectrum. For HE, spectra were constructed by the individual detector, since each detector has its own response file. We fitted those spectra with the same model parameters, but different constant coefficients.

The phase-resolved spectrum of the Crab pulsar was generated. We divided the phase of 11--250\,keV into 100 bins. The spectra of photons in each bin were fitted by a simple {\it powerlaw} with the interstellar absorption models {\it wabs}, where $N_H$ was frozen at 0.36 to fit. The background of each spectrum was the spectra of photons with the phase between 0.6 and 0.8, with the exposure normalized by the proportion of resolved phase bins to background phase bins. 

\section{results}\label{sec:results}
\subsection{pulse profile}\label{subsec:timing}
For profiles with energy lower than 100 keV, the fluxes of MPs are larger than thoses of SPs, while the fluxes of SPs are larger than those of MPs in the energy range 100--250\,keV. Figure. \ref{fig:profiles_energy} depicts a dramatic change in the different components with energy. We analyzed the evolution of the peak flux ratio($R_f$) of SP over MP, the separation($\Phi$) of two peaks in phase, and FWHM($W_1$, $W_2$) for MP and SP. 

\begin{figure}[!thb]
	\centering
	\includegraphics[width=1.0\textwidth]{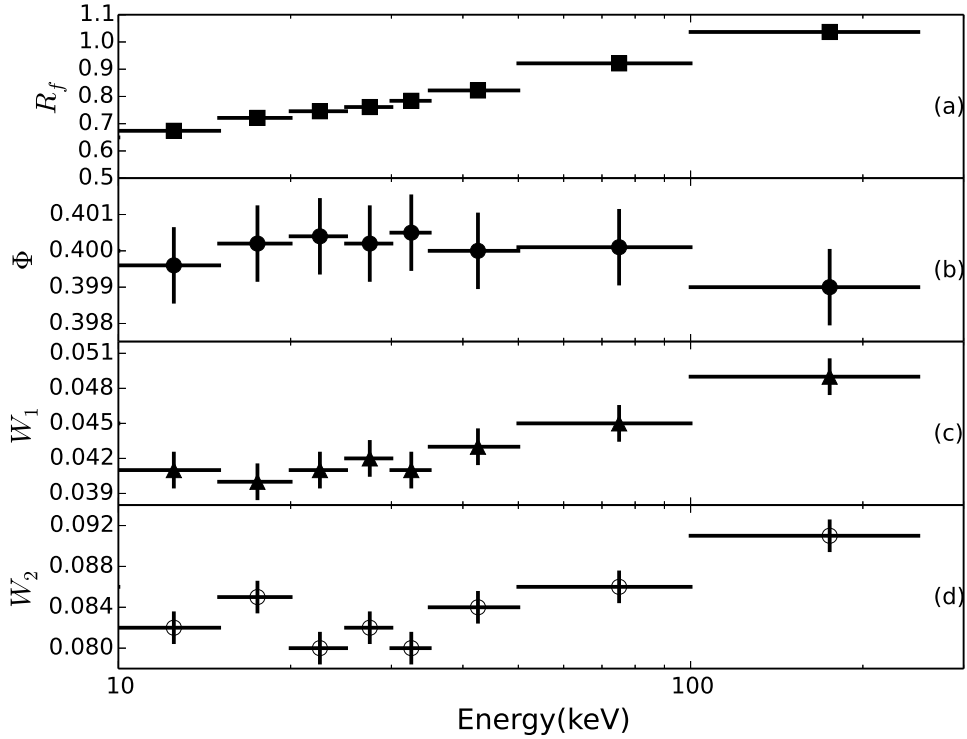}\caption{The parameters of pulse profiles evolve with energy. The panel (a) is the the flux ratio($R_f$) of the second peak over the main peak of the Crab pulsar as a function of energy. There is a increasing $R_f$ with increasing energy in the energy range 11 keV to 250 keV. In panel (b), the separation($\Phi$) of two fitted peaks of the Crab pulsar has not significant trend in energy. In panel (c) and (d), the FWHM of the MP and SP ($W_1$, $W_2$) are interpreted respectively. In energy range \SI{11}{\keV}--\SI{250}{\keV}, there are increasing trend for $W_1$ and $W_2$.\label{fig:parameters_all}}
\end{figure}

 The results are displayed in Figure. \ref{fig:parameters_all}. In panel (a), the flux ratio of the SP over the MP increases with increasing energy. The value of $R_f$ is in a range of 0.57 to 1.04 from energy \SI{11}{\keV} to \SI{250}{\keV}. In this paper, $R_f$ is larger than 1 when the energy is higher than \SI{100}{\keV}. 
 
  In panel (b), there is not a significant trend of $\Phi$ with increasing energy. The errors of $\Phi$ were calculated by the Monte Carlo method. One-hundred simulated profiles were created by sampling from the original profile under the assumption that the photon counts in every phase bin follow a Poisson distribution. Then we found the position of maximum for each peak and calculated the separation $\Phi$. Since 100 separations were obtained, errors were regarded as the 1$\sigma$ width of the Gaussian distribution. As suggested by previous work \citep{ge2012x}, a decreasing trend is expected. However, the variations of $\Phi$ in our results are about 1/1000 of the phase of one period, while the statistical errors of $\Phi$ are also at a level of 1/1000 of the phase of one period. Although the last point above \SI{100}{\keV} drops by 0.001, the trend is not significant enough to give a conclusion.
  
  $W_1$ and $W_2$ could be calculated by fitting profiles as well. In panels (c) and (d), $W_1$ and $W_2$ in the energy range 11 keV to 250 keV are plotted. 
 The statistical errors of $W_1$ and $W_2$ are given by the uncertainties in phases that propagated from the uncertainties in the parameter $\phi_0$. In the energy range 11\,keV to 250\,keV, both $W_1$ and $W_2$ have an increasing trend with increasing energy.

\subsection{Spectrum}\label{subsec:spec}

\subsubsection{Phase-averaged Spectrum}\label{subsubsec:averaged}

\begin{figure}[!htb]
\centering
\includegraphics[angle=-90, width=1.0\textwidth]{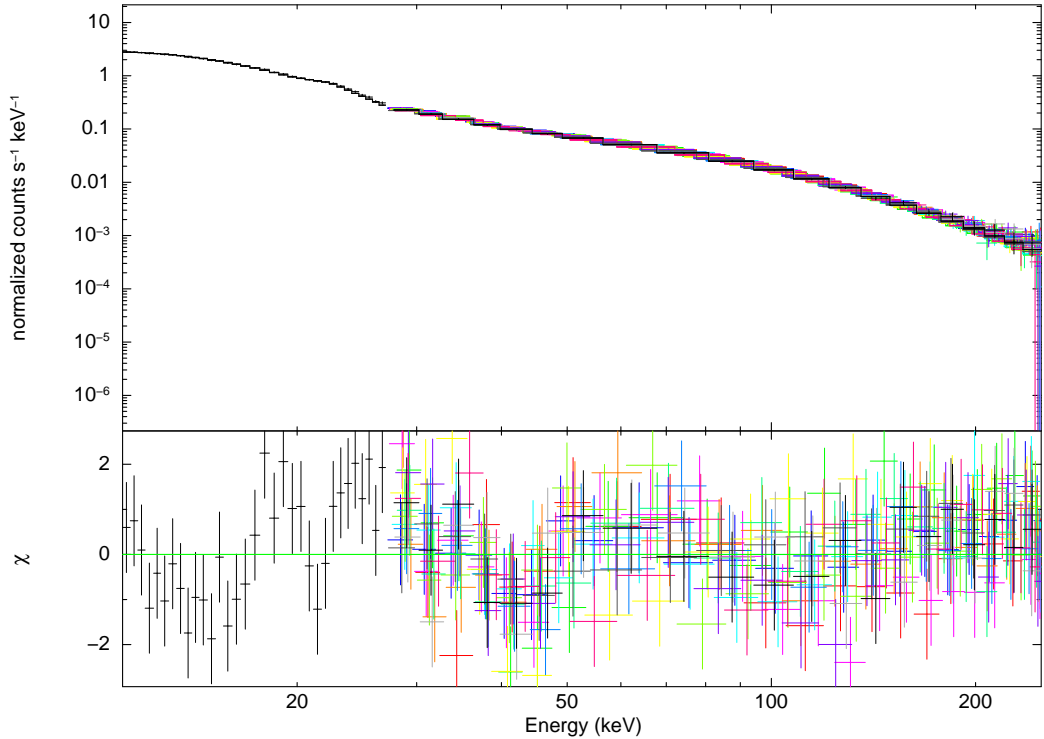}
\caption{The results of ME and 17 detectors of HE in energy range 11 keV to 250 keV. The phase-averaged spectrum of the Crab pulsar, was fitted by a logarithm power-law model, with an absorption $N_H = 0.36 \times 10^{22}~ \mathrm{cm}^{-2}$ and $\alpha=1.515$, $\beta=0.138$, $E_0$ was 1 keV respectively. The reduced $\chi^2$ is 0.959 (d.o.f 342).}\label{fig:fitting}
\end{figure}

A phase-averaged spectrum of the Crab pulsar in energy band 11\,keV to 250\,keV could be fitted with a log-parabola model. The best fitting results were an absorption models with $N_H = 0.36 \times 10^{22}\,\mathrm{cm}^{-2}$, and a log-parabola model with $\alpha=1.515$ and $\beta=0.138$, as shown in Figure. \ref{fig:fitting}. The reduced $\chi^2$ was 0.959 with 342 degrees of freedom (d.o.f). The $\chi^2$ between observed data and the model of best fitting results has no significant lines or structures.

\subsubsection{Phase-resolved Spectrum}\label{subsubsec:resolved}

Since the profiles of the Crab pulsar were obtained, we divided the phase into 100 sections for HE and ME data. A {\it powerlaw} model was employed to fit the spectra with an energy range $11\,\mathrm{keV}$--$250\,\mathrm{keV}$. 

\begin{figure}[!hbt]
\centering
\includegraphics[width=1.0\textwidth]{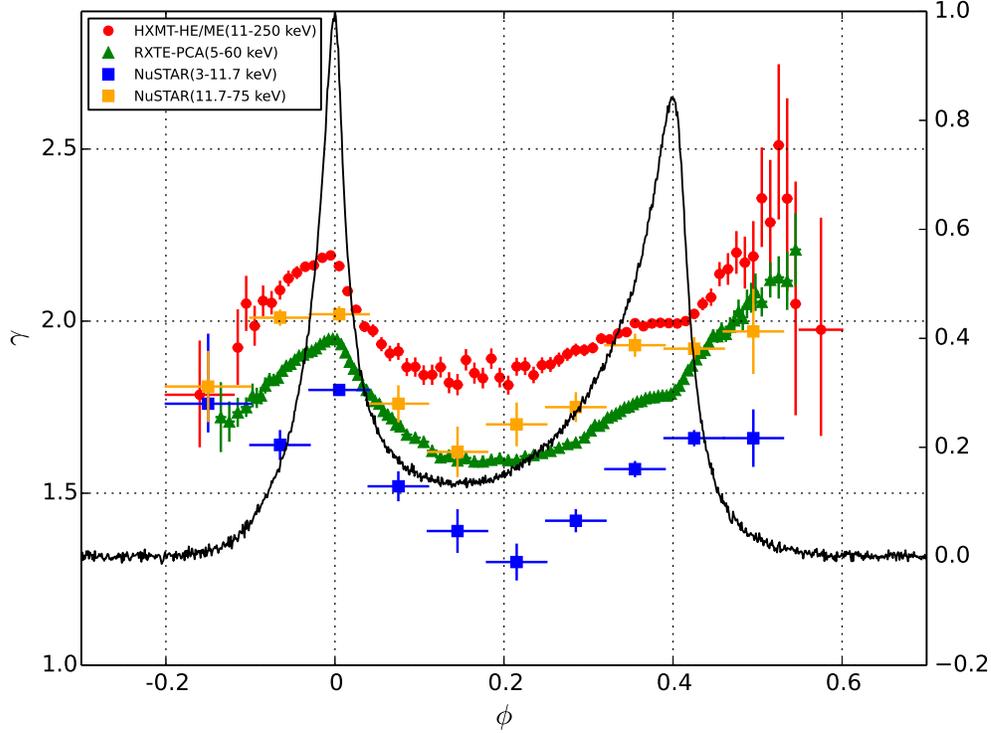}
\caption{The photon index evolution in phases. The pulse profile folded by HXMT-HE data is plotted in black for reference. The red dots represent photon indices of HE together with ME fitted by a power-law model in \si{11}-- \SI{250}{\keV}. Green triangles are photon indices of RXTE-PCA(5-60 keV). And blue squares are results of NuSTAR observation in $17''$ extraction regions in the energy range 3 keV to 11.7 keV where break energy of bknpower locates. And yellow squares are NuSTAR results above 11.7\,keV.\label{fig:gamma}}
\end{figure}

The phase-resolved spectra in energy range \SI{11}{\keV} to \SI{250}{\keV} were fitted for comparison with results of other missions. Detailed fitting results are listed in Table \ref{table:hemefitting}. The reduced $\chi^2$ of all spectra are between 0.94 -- 1.24. Those spectra near the background phases(0.6--0.8) are relatively poorly fitted with large error bars ascribed to low statistical significance. As dipicted in Figure.\ref{fig:gamma}, the photon indices increase from the edge of the background phase to the maximum position of the MP(phase -0.2 to phase 0). From the tail of the MP to the bridge region, the photon indices gradually decrease. 

 From the maximum position of the SP to the edge of the background phase(phase 0.4 to phase 0.6), the photon indices increased more steeply.

\section{Conclusions}\label{sec:disscussion}
The results of timing and phase-resolved spectroscopy suggested good performance of \insight{}. The microsecond time resolution of \insight{} and good timing residuals of {\it HXMT} data yielded light curves and profiles in 11--250\,keV for the Crab pulsar.

 The increasing trend of $R_f$ with energy is consistent with what previous work has demonstrated \citep{Kuiper2001}, in which the flux ratio increases monotonically with increasing energy from around \SI{1}{\eV} to around \SI{1}{\MeV}. \cite{Kuiper2001} focused on the integrated flux of each peak, phase intervals of which are defined by \cite{fierro1998phase}. We analyzed the peak flux of MP and SP instead. However, the trends are  similar to each other. A significant increase in the ratio of SP over MP above $11\,\mathrm{keV}$ was present in the {\it HXMT} energy band, which is consistent with previous work\citep{Tompkins1997, Eikenberry1997}.

 Our results suggest a dramatic change in profile with energy. It is also consistent with the previous results \citep{ge2012x}. There is no significant trend for the separation between two peaks. This implies that the emitting regions from different energy ranges have no significant variance or it is not significant until more observations and data are used. A measurement of FWHM from two peaks was carried out in the energy range 11--250\,keV as well. There is an increasing trend of FWHM for the two peaks. 

 In Figure. \ref{fig:gamma}, we compared the evolution of our photon indices in phase with the results of {\it RXTE}\citep{ge2012x} and {\it NuSTAR}\citep{madsen2015broadband}. The phase-resolved spectroscopy results of \insight{} confirmed a homogenous trend in different energy bands and an evolution in energy. In the outer gap model or slot gap model, the X-ray photons of the Crab come from synchrotron radiation of secondary pairs. In the phase-resolved spectra, the spectral indices follow a similar trend with the hardest spectra occurring in the bridge region and the softer spectra located in the peak. With broadband observation of \insight{}, the photon indices that evolve as a function of energy were contributing to understanding the pair cascade processes in a pulsar magnetosphere. The broadband X-ray observations share a common property. We expect a more detailed model constraint for this broadband phenomenon. 


\begin{acknowledgements}
This work is supported by the 
National Key R\&D Program of China (2016YFA0400800) 
and the National Natural Science Foundation of China under 
grants 11503027, 11673023, U1838201 and U1838104. 
This work made use of the data from the HXMT mission, 
a project funded by China National Space Administration (CNSA) 
and the Chinese Academy of Sciences (CAS). And we thank the suggestion from Bing Li.

\end{acknowledgements}

\bibliography{msRAA-2019-0034R1_bibtex.bib}
\bibliographystyle{raa}

\begin{longtable}{cccc}
\hline
\endfoot
\caption{The phase-resolved spectral analysis results of HE together with those of ME. A {\it powerlaw} model was used. \label{table:hemefitting}}
\\\hline

Pulse Phase Range &  Normalization & Spectral Index   & Reduced $\chi^2$                                                    \\
\hline
\endhead
0.0 -- 0.01 & $6.42_{-0.09}^{+0.09}$ & $2.16_{-0.005}^{+0.005}$ & 1.197 \\
0.01 -- 0.02 & $6.77_{-0.13}^{+0.13}$ & $2.087_{-0.007}^{+0.006}$ & 1.139 \\
0.02 -- 0.03 & $4.98_{-0.12}^{+0.12}$ & $2.033_{-0.009}^{+0.008}$ & 1.117 \\
0.03 -- 0.04 & $2.67_{-0.08}^{+0.09}$ & $1.984_{-0.012}^{+0.012}$ & 0.976 \\
0.04 -- 0.05 & $1.75_{-0.07}^{+0.07}$ & $1.972_{-0.014}^{+0.015}$ & 1.013 \\
0.05 -- 0.06 & $1.18_{-0.05}^{+0.05}$ & $1.933_{-0.017}^{+0.016}$ & 1.003 \\
0.06 -- 0.07 & $0.87_{-0.04}^{+0.05}$ & $1.906_{-0.018}^{+0.019}$ & 1.035 \\
0.07 -- 0.08 & $0.73_{-0.04}^{+0.04}$ & $1.912_{-0.020}^{+0.021}$ & 0.990 \\
0.08 -- 0.09 & $0.57_{-0.03}^{+0.04}$ & $1.866_{-0.022}^{+0.022}$ & 0.997 \\
0.09 -- 0.1 & $0.49_{-0.03}^{+0.03}$ & $1.867_{-0.024}^{+0.024}$ & 0.992 \\
0.1 -- 0.11 & $0.42_{-0.03}^{+0.03}$ & $1.843_{-0.024}^{+0.025}$ & 1.058 \\
0.11 -- 0.12 & $0.39_{-0.03}^{+0.03}$ & $1.844_{-0.026}^{+0.026}$ & 1.047 \\
0.12 -- 0.13 & $0.39_{-0.03}^{+0.03}$ & $1.866_{-0.026}^{+0.026}$ & 0.969 \\
0.13 -- 0.14 & $0.34_{-0.02}^{+0.02}$ & $1.821_{-0.026}^{+0.027}$ & 0.993 \\
0.14 -- 0.15 & $0.31_{-0.02}^{+0.02}$ & $1.815_{-0.026}^{+0.028}$ & 0.965 \\
0.15 -- 0.16 & $0.38_{-0.03}^{+0.03}$ & $1.887_{-0.026}^{+0.028}$ & 1.065 \\
0.16 -- 0.17 & $0.33_{-0.02}^{+0.02}$ & $1.849_{-0.026}^{+0.026}$ & 0.987 \\
0.17 -- 0.18 & $0.33_{-0.02}^{+0.02}$ & $1.834_{-0.026}^{+0.026}$ & 1.008 \\
0.18 -- 0.19 & $0.38_{-0.03}^{+0.03}$ & $1.891_{-0.026}^{+0.026}$ & 1.037 \\
0.19 -- 0.2 & $0.33_{-0.02}^{+0.02}$ & $1.836_{-0.025}^{+0.024}$ & 1.015 \\
0.2 -- 0.21 & $0.34_{-0.02}^{+0.02}$ & $1.814_{-0.024}^{+0.024}$ & 1.010 \\
0.21 -- 0.22 & $0.41_{-0.02}^{+0.03}$ & $1.868_{-0.023}^{+0.021}$ & 0.978 \\
0.22 -- 0.23 & $0.42_{-0.02}^{+0.03}$ & $1.869_{-0.022}^{+0.022}$ & 1.001 \\
0.23 -- 0.24 & $0.43_{-0.02}^{+0.02}$ & $1.843_{-0.020}^{+0.021}$ & 1.048 \\
0.24 -- 0.25 & $0.48_{-0.02}^{+0.02}$ & $1.872_{-0.020}^{+0.018}$ & 0.975 \\
0.25 -- 0.26 & $0.51_{-0.02}^{+0.03}$ & $1.874_{-0.018}^{+0.017}$ & 0.997 \\
0.26 -- 0.27 & $0.59_{-0.03}^{+0.03}$ & $1.888_{-0.016}^{+0.017}$ & 0.981 \\
0.27 -- 0.28 & $0.66_{-0.03}^{+0.03}$ & $1.904_{-0.016}^{+0.015}$ & 1.008 \\
0.28 -- 0.29 & $0.75_{-0.03}^{+0.03}$ & $1.916_{-0.014}^{+0.015}$ & 1.010 \\
0.29 -- 0.3 & $0.79_{-0.03}^{+0.03}$ & $1.916_{-0.014}^{+0.013}$ & 1.057 \\
0.3 -- 0.31 & $0.91_{-0.03}^{+0.03}$ & $1.922_{-0.012}^{+0.013}$ & 0.989 \\
0.31 -- 0.32 & $1.07_{-0.03}^{+0.04}$ & $1.949_{-0.012}^{+0.012}$ & 1.093 \\
0.32 -- 0.33 & $1.20_{-0.03}^{+0.04}$ & $1.947_{-0.010}^{+0.010}$ & 1.086 \\
0.33 -- 0.34 & $1.38_{-0.04}^{+0.04}$ & $1.964_{-0.010}^{+0.010}$ & 1.110 \\
0.34 -- 0.35 & $1.58_{-0.04}^{+0.04}$ & $1.968_{-0.008}^{+0.009}$ & 1.053 \\
0.35 -- 0.36 & $1.88_{-0.04}^{+0.04}$ & $1.994_{-0.008}^{+0.007}$ & 1.105 \\
0.36 -- 0.37 & $2.06_{-0.04}^{+0.04}$ & $1.985_{-0.007}^{+0.008}$ & 1.084 \\
0.37 -- 0.38 & $2.35_{-0.04}^{+0.04}$ & $1.993_{-0.007}^{+0.006}$ & 1.150 \\
0.38 -- 0.39 & $2.70_{-0.04}^{+0.04}$ & $1.995_{-0.006}^{+0.005}$ & 1.151 \\
0.39 -- 0.4 & $3.03_{-0.05}^{+0.04}$ & $1.994_{-0.004}^{+0.006}$ & 1.177 \\
0.4 -- 0.41 & $3.44_{-0.05}^{+0.05}$ & $1.993_{-0.006}^{+0.004}$ & 1.195 \\
0.41 -- 0.42 & $3.82_{-0.07}^{+0.07}$ & $2.0_{-0.006}^{+0.007}$ & 1.206 \\
0.42 -- 0.43 & $3.82_{-0.10}^{+0.11}$ & $2.021_{-0.009}^{+0.010}$ & 1.150 \\
0.43 -- 0.44 & $3.02_{-0.12}^{+0.12}$ & $2.05_{-0.014}^{+0.015}$ & 1.057 \\
0.44 -- 0.45 & $2.06_{-0.12}^{+0.12}$ & $2.069_{-0.020}^{+0.022}$ & 1.029 \\
0.45 -- 0.46 & $1.66_{-0.13}^{+0.15}$ & $2.137_{-0.032}^{+0.031}$ & 1.025 \\
0.46 -- 0.47 & $1.26_{-0.14}^{+0.16}$ & $2.151_{-0.041}^{+0.043}$ & 1.016 \\
0.47 -- 0.48 & $1.03_{-0.15}^{+0.18}$ & $2.199_{-0.057}^{+0.059}$ & 1.006 \\
0.48 -- 0.49 & $0.70_{-0.12}^{+0.15}$ & $2.17_{-0.072}^{+0.072}$ & 0.997 \\
0.49 -- 0.5 & $0.58_{-0.13}^{+0.18}$ & $2.188_{-0.096}^{+0.098}$ & 1.058 \\
0.5 -- 0.51 & $0.68_{-0.21}^{+0.32}$ & $2.357_{-0.138}^{+0.143}$ & 0.970 \\
0.51 -- 0.52 & $0.45_{-0.16}^{+0.27}$ & $2.287_{-0.167}^{+0.177}$ & 0.950 \\
0.52 -- 0.53 & $0.59_{-0.26}^{+0.52}$ & $2.511_{-0.212}^{+0.231}$ & 1.000 \\
0.53 -- 0.54 & $0.35_{-0.18}^{+0.41}$ & $2.356_{-0.264}^{+0.288}$ & 1.021 \\
0.54 -- 0.55 & $0.09_{-0.05}^{+0.16}$ & $2.05_{-0.319}^{+0.352}$ & 0.999 \\
0.55 -- 0.6 & $0.04_{-0.02}^{+0.06}$ & $1.975_{-0.305}^{+0.322}$ & 0.964 \\
0.8 -- 0.88 & $0.01_{-0.00}^{+0.00}$ & $1.786_{-0.149}^{+0.153}$ & 0.953 \\
0.88 -- 0.89 & $0.05_{-0.01}^{+0.02}$ & $1.923_{-0.104}^{+0.108}$ & 0.988 \\
0.89 -- 0.9 & $0.14_{-0.03}^{+0.04}$ & $2.051_{-0.076}^{+0.076}$ & 0.943 \\
0.9 -- 0.91 & $0.16_{-0.02}^{+0.03}$ & $1.986_{-0.054}^{+0.054}$ & 0.961 \\
0.91 -- 0.92 & $0.30_{-0.03}^{+0.04}$ & $2.059_{-0.041}^{+0.040}$ & 1.042 \\
0.92 -- 0.93 & $0.44_{-0.04}^{+0.04}$ & $2.053_{-0.029}^{+0.031}$ & 1.011 \\
0.93 -- 0.94 & $0.68_{-0.04}^{+0.05}$ & $2.09_{-0.024}^{+0.024}$ & 0.978 \\
0.94 -- 0.95 & $1.00_{-0.05}^{+0.06}$ & $2.124_{-0.019}^{+0.018}$ & 1.023 \\
0.95 -- 0.96 & $1.39_{-0.06}^{+0.06}$ & $2.141_{-0.015}^{+0.015}$ & 1.030 \\
0.96 -- 0.97 & $1.95_{-0.06}^{+0.06}$ & $2.158_{-0.011}^{+0.012}$ & 1.046 \\
0.97 -- 0.98 & $2.59_{-0.06}^{+0.07}$ & $2.162_{-0.008}^{+0.008}$ & 1.124 \\
0.98 -- 0.99 & $3.67_{-0.07}^{+0.07}$ & $2.184_{-0.007}^{+0.006}$ & 1.138 \\
0.99 -- 1.0 & $5.03_{-0.08}^{+0.08}$ & $2.191_{-0.004}^{+0.006}$ & 1.244 \\
\end{longtable}

\end{document}